\documentclass[nofootinbib,aps,amsmath,amssymb,a4paper,prd,twocolumn]{revtex4}

\usepackage{graphicx}
\usepackage{color}

\usepackage{units}

\allowdisplaybreaks

\usepackage{hyperref}
\hypersetup{
   pdfnewwindow=true,     
   colorlinks=true,       
   linkcolor=black,       
   citecolor=blue,        
   filecolor=black,       
   urlcolor=blue          
}
\usepackage{amsmath} 
\usepackage{amsthm} 
\usepackage{amssymb}	
 
\newcommand{\matrixel}[3]{\left< #1 \vphantom{#2#3} \right|
 #2 \left| #3 \vphantom{#1#2} \right>} 
\let\baraccent=\= 
\renewcommand{\=}[1]{\stackrel{#1}{=}} 

\theoremstyle{definition}

\theoremstyle{remark}

\usepackage{bbm}

\begin{document}
\title{\Large{\textit{{\bf{Renormalizable SU(5) Unification}}}}}
\author{Pavel \surname{Fileviez Perez}}
\author{Clara \surname{Murgui}}
\affiliation{\\ \small{
\vspace{0.3cm}
Particle and Astroparticle Physics Division,\\
Max-Planck-Institut f\"ur Kernphysik,\\
Saupfercheckweg 1, 69117 Heidelberg, Germany}
}
\begin{abstract}
We propose a simple renormalizable grand unified theory based on the $SU(5)$ gauge symmetry where the neutrino masses 
are generated at the quantum level through the Zee mechanism. In this model the same Higgs needed to correct the mass relation between charged 
leptons and down-type quarks plays a crucial role to generate neutrino masses. We show that in this model one can 
satisfy the constrains coming from the unification of gauge couplings and the mechanism for neutrino masses is discussed in detail. 
The predictions for proton decay are discussed in order to understand the testability at current and future experiments such as Hyper-Kamiokande.
This simple theory predicts a light colored octet which could give rise to exotic signatures at the LHC.
\end{abstract}

\maketitle

\section{Introduction}
\label{sec:introduction}
The idea of grand unification has been considered as one of the appealing ways to think about physics beyond the Standard Model.
The simplest grand unified theory was proposed in 1974 by H.~Georgi and S.~Glashow~\cite{Georgi:1974sy}. 
This theory is based on the $SU(5)$ gauge symmetry and makes several striking predictions.  
It predicts a new phenomena in nature, the decay of the proton~\cite{Pati}. For a review see Ref.~\cite{Nath:2006ut}. In this context the Standard Model interactions can be seen as different 
manifestations of the same fundamental interaction at the low scale. One can understand the electric charge quantization 
and predict the Weinberg's angle.  Last, but not least, this theory predicts the existence of a great desert between the electroweak scale 
and unification scale where the $SU(5)$ symmetry could be realised.   

It is well-known that the Georgi-Glashow model~\cite{Georgi:1974sy} is ruled out by the experiments.
In this context the Standard Model matter fields are unified in the $\overline{5}=(\ell, \ d^c)=(1,2,-1/2)\oplus(\bar{3},1,1/3)$ and $10=(q, \ u^c, \ e^c)=(3,2,1/6)\oplus (\bar{3},1,-2/3)\oplus (1,1,1)$ 
representations, while the minimal Higgs sector is composed of the $5_H=(T, \ H_1)$ and $24_H = (\Sigma_8, \Sigma_{3}, \Sigma_{(3,2)}, \Sigma_{(\bar{3},2)},\Sigma_{24})$ Higgses. This model is ruled 
out because one cannot reproduce the experimental values of the gauge couplings measured at the low scale. 

In this letter we discuss the different realistic scenarios for $SU(5)$ unification proposed in Refs.~\cite{Dorsner:2005fq,Dorsner:2005ii,Dorsner:2006hw,Bajc:2006ia,Bajc:2007zf,Dorsner:2006fx}.
After a detailed discussion we propose a new grand unified theory where the charged fermion masses are generated at the renormalizable level, while the neutrino masses are generated 
at the quantum level through the Zee mechanism~\cite{Zee:1980ai}. For the first time, in the context of a grand unified theory based on the $SU(5)$ symmetry, one can find a simple relation 
between charged fermion masses and neutrino masses. We investigate the unification of gauge interactions and show we can have consistent scenarios for unification when 
the colored octet present in the theory is not very heavy. Therefore, this theory predicts the possibility to observe exotic signals at current and future colliders.
The predictions for proton decay are shown in order to understand the testability of the theory at the Super-Kamiokande or Hyper-Kamiokande experiments. 
The model proposed in this letter can be considered as one of the simplest renormalizable $SU(5)$ models.

\section{Realistic $SU(5)$ Theories}
It is possible to write down very simple realistic extensions of the Georgi-Glashow model. This issue has 
been investigated by many experts in the field. Here we discuss the two simple models:

\begin{itemize}

\item Type II-SU(5): In this case one generates neutrino masses through the Type II~\cite{TypeII} seesaw mechanism and making 
use of the higher-dimensional operators one can have a consistent relation between the masses for down quarks and charged leptons.
In this model the unification constraints imply that the leptoquark with quantum numbers $(3,2,1/6)$ living in the $15_H=(3,2,1/6)\oplus (1,3,1) \oplus (6,1,-2/3)$ must be light. 
This non-renormalizable model has been studied in detail in Refs.~\cite{Dorsner:2005fq,Dorsner:2005ii,Dorsner:2006hw}.

\item Type III-SU(5): Using a combination of the Type I~\cite{TypeI} and Type III~\cite{TypeIII} seesaw mechanisms adding the $24$ fermionic representation 
one can define a simple realistic model based on $SU(5)$. See Refs.~\cite{Bajc:2006ia,Bajc:2007zf,Dorsner:2006fx} for a detailed study.
This model is non-renormalizable since higher-dimensional operators are used to correct the mass relation between the down quarks and charged leptons, 
and to understand the splitting between the different fields in the $24$ representation. The unification constraints in this context tell us that 
the fermionic triplet needed for Type III seesaw must be light. 

\end{itemize}

The models discussed above are non-renormalizable. It is very well-known that in order to have a realistic relation between down quarks and charged leptons 
at the renormalizable level one must add the $45_H$ representation~\cite{Georgi:1979df}.  Now, one could ask: {\textit{What is the simplest realistic renormalizable model 
based on $SU(5)$ ?}}. 

The scenarios discussed above can be used as a guide to define the simplest model. In the Type II-SU(5) case one has two extra Higgses, $15_H$ and $45_H$~\cite{Dorsner:2007fy}, or in the Type III-SU(5) case one has $45_H$ and one extra fermionic $24$~\cite{Perez:2007rm,Perez:2008ry} representation. 
A different case could be considered where the neutrino masses are generated through Type I seesaw and one has at least two extra singlets, right-handed neutrinos, and the extra Higgs $45_H$. This scenario can be realistic~\cite{Dorsner:2006dj} but one expects naively that the fermionic singlets should get mass from above GUT scale since their masses are not protected by the SU(5) gauge symmetry. 

In the next section we will show that a realistic renormalizable model can be defined with two extra Higgses, $10_H$ and $45_H$, where neutrino masses 
are generated through the Zee mechanism.  Here the $45_H$ plays also a crucial role to generate neutrino masses. 
%
\section{Zee model and $SU(5)$ Unification}
\label{sec:zee_model}
%
In the Zee model~\cite{Zee:1980ai} for neutrino masses two extra Higgses are needed to generate neutrino masses at one-loop level.
One has two Higgs doublets including the SM Higgs boson. In our notation we have $H_a \sim (1,2,1/2)$ where $a=1,2$ 
and $\delta^+ \sim (1,1,1)$. The relevant interactions are given by
\begin{equation}
V_{\text{Zee}}= \ell_{L}  \lambda \ \ell_{L}  \delta^{+}  \ + \  \overline{\ell_L}  Y_a  H_a  e_R \ + \  \mu H_1 H_2 \delta^- \ + \ \text{h.c.}, 
\end{equation} 
where $\ell_L \sim (1,2,-1/2)$, $e_R \sim (1,1,-1)$, the matrix $\lambda$ is antisymmetric in the flavor space, and $Y_a$ are the Yukawa matrices for the two Higgses present in the theory.
Here the global $B-L$ symmetry is broken due to the simultaneous presence of the first Yukawa interaction  proportional to $\lambda$ and the last term proportional to the $\mu$ parameter.  
%
%
\begin{widetext}
Computing the neutrino mass matrix in the broken phase one finds
\begin{eqnarray}
M_{\nu} &=& \frac{1}{8 \pi^2} \left( \lambda M_{e} \left(  Y_1^\dagger \cos \beta - Y_2^\dagger \sin \beta \right) +  \left(  Y_1^* \cos \beta - Y_2^* \sin \beta \right) M_e^T \lambda^T \right) \sin 2 \theta_+ \rm{Log} \left( \frac{m^2_{h_2^+}}{m^2_{h_1^+}} \right).
\end{eqnarray}   
\end{widetext}
where the angle $\beta$ defines the relation between the charged Higgses in the Higgs doublets
\begin{eqnarray}
H_1^\pm &=& \cos \beta H^\pm + \sin \beta G^\pm, \\
H^\pm_2 &=& - \sin \beta H^\pm + \cos \beta G^\pm.
\end{eqnarray}
The mixing angle $\beta$ is defined by the diagonalization of the mass matrix for the $H_i^{\pm}$ Higgses and one finds $\tan \beta = v_1/v_2$. Here $v_1$ and $v_2$ are the vacuum expectation values of the neutral Higgses satisfying the relation $v_1^2 + v_2^2=v^2$ where $v=246$ GeV. 
Here $G^{\pm}$ are the Goldstone bosons eaten up by the $W^{\pm}$. The mixing $\theta_+$ defines the mixing between the $H^+$ and the singlet $\delta^+$
\begin{eqnarray}
\delta^\pm &=& \cos \theta_+ h_1^\pm + \sin \theta_+ h_2^\pm, \\
H^\pm &=& - \sin \theta_+ h_1^\pm + \cos \theta_+ h_2^\pm.
\end{eqnarray}
In the above equation $\theta_+$ is proportial to the $\mu$ parameter.
Therefore, the neutrino masses are also proportional to the $\mu$ parameter which breaks the global $B-L$. Then, when $\mu$ is small one can naturally have small neutrino masses 
and this parameter is protected by the symmetry.  
Notice that when we assume the Zee-Wolfenstein model~\cite{Zee:1980ai,Wolfenstein:1980sy} where only one Higgs couples to leptons the resulting mass matrix has zero diagonal entries.
This particular scenario is ruled out by the experiments~\cite{Smirnov:1996bv,He:2003ih}. However, in the general Zee model for neutrino masses one has enough freedom 
to reproduce the values for neutrino mixings and masses. See for example~\cite{He:2011hs} for a recent study of the Zee model.

Now, the Zee model can be realized in a grand unified theory based on $SU(5)$ adding the new Higgs in the antisymmetric representation $10_H$ which contains the charged singlet $\delta^+$~\cite{Zee:1980ai}.
A second Higgs doublet is needed as we discussed. However, if we are in the minimal renormalizable $SU(5)$ there is no need to introduce a new Higgs because the $45_H$ representation contains already a second Higgs doublet.
This is a crucial observation which allow us to define a simple model where the neutrino masses are related to the charged fermion masses.
As we know, the $45_H$ representation has to be in the renormalizable theory in order to correct the mass relation between the charged leptons and down quarks. In the $SU(5)$ language the needed interactions for the Zee model read as 
\begin{eqnarray}
V_{SU(5)} &\supset & \lambda \bar{5}\, \bar{5}\,10_H  + \bar{5}\,10 \left( Y_1^* 5^*_H - \frac{1}{6}Y_2^* 45^*_H \right) \nonumber \\
&-& \frac{1}{6} \mu \ 5_H 45_H 10_H^* \ + \ \text{h.c.}
\end{eqnarray}    
In this way we can see that the minimal renormalizable $SU(5)$ without extra singlets is defined by the Zee mechanism for neutrino masses. 
Clearly, one can see that since $Y_1$ and $Y_2$ cannot be diagonalized simultaneously, the diagonal elements of the 
neutrino mass matrix are not zero even if $\lambda$ is antisymmetric. Therefore, the model has enough freedom to be 
consistent with the experimental values for neutrino masses and mixings. Using the relation between the charged fermion masses and the Yukawa couplings in the renormalizable theory
\begin{eqnarray}
  \sqrt{2} M_d &=&Y_1^T v_5 - \frac{1}{3} Y_2^T v_{45},\\
  \sqrt{2} M_e &=& Y_1 v_5  + Y_2 v_{45},
 \label{masses_yukawes}
\end{eqnarray}
the neutrino mass matrix can be written as
%
\begin{eqnarray}
M_{\nu} &=& \lambda M_{e} \left(  c_e M_e^\dagger + 3 c_d  M_d^* \right) \nonumber  \\
&+ & \left( c_e M_e^* + 3 c_d M_d^\dagger \right) M_e^T \lambda^T. 
\end{eqnarray}   
%
where the coefficients $c_e$ and $c_d$ encode the information of the 1-loop radiative correction to the neutrino masses and are given in 
the Appendix.   
This relation is quite interesting because in $SU(5)$ a relation between the neutrino masses and the charged fermion masses is not expected.
Notice that the anti-symmetric matrix $\lambda$ defines this relation and it has only three free parameters. This is one of our main results.

Working in the basis where $M_e$ and $M_u$ are diagonal, using $M_d=D_c^* M_d^{diag} V_{CKM}^\dagger$, and neglecting all phases for simplicity
 one finds
\begin{eqnarray}
M_\nu &=& \lambda M_e^{diag} \left( c_e M_e^{diag}  + 3 c_d  D_c M_d^{diag} V_{CKM}^T\right) \nonumber \\
&+&  \left( c_e M_e^{diag} + 3 c_d V_{CKM} M_d^{diag} D_c^T \right) M_e^{diag} \lambda^T. \nonumber \\
\end{eqnarray}
Here $D_c$ is the matrix which rotates the $d^c$ quarks, which has only three parameters in the real case, and the coefficients $c_e$ and $c_d$ are given in the Appendix.
As one can see the above matrix has enough parameters to reproduce the values for the mixing angles and masses.
Notice that in this model the ratio between the fermion masses cannot be predicted but one can have a simple relation between 
them which can be constrained using the experimental values for fermion masses and mixings. In a future publication we will 
investigate the constrains on the free parameters including the running of the fermion masses and possible predictions for lepton flavor violating processes. 
%
\subsection{Predictions from Unification}
\label{sec:unification}
In this section we show the simplest scenarios where one can have unification of the gauge interactions in agreement with the experiments.
The renormalization group equations for the gauge couplings can be written as
\begin{equation}
 \displaystyle \alpha_i^{-1}(M_Z) = \alpha^{-1}_{GUT}+{ B_i \over 2\pi} \text{Log}\left({M_{GUT} \over M_Z}\right),
 \label{RGE_coupling}
\end{equation}
where
\begin{equation}
B_i=b_i^{SM} + b_{iI} r_I, \ \textrm{and} \ r_I= \frac{\text{Log} {(M_{GUT} / M_I)}}{\text{Log} {(M_{GUT} / M_Z)}}.
\end{equation}
Here $M_I$ is the mass of the particle living in the great desert. The equations for the running of the gauge couplings can be rewritten in a more suitable form in terms of the differences in the coefficients 
$B_{ij}=B_i-B_j$ and low energy observables~\cite{Giveon:1991zm} at the electroweak scale. These equations read as
\begin{eqnarray}
 \displaystyle {B_{23}\over B_{12}}&=&{5\over 8}\left({\text{sin}^2\theta_W (M_Z) - \alpha (M_Z) / \alpha_s (M_Z) \over 3/8-\text{sin}^2\theta_W (M_Z)}\right), \nonumber \\
 \label{ratio_unification}\\
 \displaystyle \text{Log}\left({M_{GUT}\over M_Z}\right)&=&{16\pi \over 5\alpha (M_Z)}\left({3/8-\text{sin}^2\theta_W (M_Z) \over B_{12}}\right). \label{gut_scale}
\end{eqnarray}
\\
Adopting the experimental values, $\alpha (M_Z)^{-1}=127.94$, $\sin^2 \theta_W (M_Z)=0.231$, and $\alpha_s (M_Z)=0.1185$~\cite{PDG} one finds 
\begin{eqnarray}
 \textrm{Log} \frac{M_{GUT}}{M_Z}  &=&{184.87 \over B_{12}}, \\  \nonumber \\
 {B_{23}\over B_{12}}&=& 0.718.
\end{eqnarray}
Therefore, these values must be reached in order to achieve unification of the couplings. 

\begin{widetext}
\begin{table}[h] 
  \vspace{0.5cm}
 \begin{tabular}{ c | c c | c c | c  c  c  c  c  c  c | c  c  c }
  \hline
 \multicolumn{3}{c}{$5_H$} & \multicolumn{2}{c}{$24_H$} &  \multicolumn{7}{c}{$45_H$} &  \multicolumn{3}{c}{$10_H$} \\
  \hline
$B_{ij}$   & $H_1$ & $T$ & $\Sigma_8$ & $\Sigma_3$  & $\Phi_1$ & $\Phi_2$ & $\Phi_3$ & $\Phi_4$ & $\Phi_5$ & $\Phi_6$ & $H_2$ & $\delta^{+}$ & $\delta_{(3,2)}$ & $\delta_{T}$\\  
  \hline
  $B_{12}$  & $-{1\over15}$ &
  ${1\over 15}r_{T}$ & 0 & $-{1\over 3}r_{\Sigma_3}$ & $-{8\over15}r_{\Phi_1}$ & ${2\over15}r_{\Phi_2}$ & $-{9\over5}r_{\Phi_3}$ & ${17\over15}r_{\Phi_4}$ & ${1\over15}r_{\Phi_5}$ &
  ${16\over15}r_{\Phi_6}$ & $-{1\over15}r_{H_2}$ & ${1\over5}r_{\delta^+}$ & $-{7\over15}r_{\delta_{(3,2)}}$ & ${4\over15}r_{\delta_T}$ \\
  $B_{23}$  & ${1\over6}$ & $-{1\over 6}r_T$ & 
  $-{1\over2}r_{\Sigma_8}$ & ${1\over 3}r_{\Sigma_3}$ & $-{2\over 3}r_{\Phi_1}$ & $-{5\over 6}r_{\Phi_2}$ & ${3\over 2}r_{\Phi_3}$ & ${1\over 6}r_{\Phi_4}$ & $-{1\over 6}r_{\Phi_5}$ &
  $-{1\over 6}r_{\Phi_6}$ & ${1\over 6}r_{H_2}$ & 0 & ${1\over6}r_{\delta_{(3,2)}}$ & $-{1\over6}r_{\delta_T}$\\ 
  \hline
  \end{tabular}
 \label{B_coef_SU5}
 \caption{$B_{ij}$ coefficients.}
 \end {table}
\end{widetext}
In Table I we show the contributions of the physical fields in $5_H$, $24_H$, $10_H$ and $45_H$ to the running of the gauge couplings. The triplet $T$ in $5_H$ has to be heavy, $M_T \gtrsim 10^{12}$ GeV, 
in order to satisfy the proton decay bounds. The only field in $24_H$ which can help for the unification is $\Sigma_3 \sim (1,3,0)$. Even if $\Sigma_3$ and $H_1$ are at the electroweak scale the constraints in Eqs.~(15) and (16) cannot be satisfied because $B_{23}^{GG}/B_{12}^{GG} \leq 0.6$. Since only the fields with negative contribution to $B_{12}$ and positive contribution to $B_{23}$ can help to achieve unification in agreement with the experiment, only the fields $\Phi_3$ and $H_2$ in $45_H$ can help. The field $\Phi_1$ also can help to increase the GUT scale and suppress proton decay. 

In the $10_H$ only the field $\delta_{(3,2)}$ could help to achieve consistent unification but one should notice that it mediates proton decay.  
The $\delta_{(3,2)}$ couples to fermions through the term $\lambda \, \bar{5} \, \bar{5} \, 10_H $ in the folowing way 
$ \lambda \epsilon_{\alpha \beta} d^c \ell^\alpha \delta_{(3,2)}^\beta$, where $\delta_{(3,2)}= (\delta_{(3,2)}^{2/3},\delta_{(3,2)}^{-1/3})$ in the 
$SU(2)_L$ space. Hence, $\delta_{(3,2)}$ alone cannot mediate proton decay. 
However, the term $-\frac{1}{3}\mu T_i^*{H_2}_{\alpha}^*\delta^{i\alpha}_{(3,2)} \in -\frac{1}{6}\mu 5_H^*45_H^*10_H$ in the scalar potential 
together with the above interaction give us a contribution to proton decay as shown in Fig.~1.
\begin{figure}[h]
\includegraphics[width=0.8\linewidth]{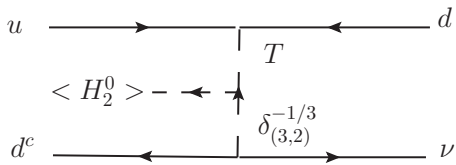}
\caption{$B-L$ violating proton decay contribution.}
\end{figure}

A qualitative study on the bounds of the delta mass scale can be made by considering the effective coupling of the process shown in Fig.~1, 
which is given by
\begin{equation}
Y_3 \frac{\lambda \, \mu \, v_2}{M_{\delta_{(3,2)}}^2M_T^2} ud\bar{\nu} \bar{d^c}. 
\end{equation}
Here $Y_3$ is the coupling in the $Y_3 10 10 5_H$ interaction.
In order to satisfy the bounds on proton decay, $\mu \, \lambda \, v_2 / M_{\delta_{(3,2)}}^2 M_T^2 \lesssim 1/(10^{12} \rm{GeV})^2$ as in the usual Higgs mediated $d=6$ proton decay contribution. 
Notice that, due to the presence of the triplet mass squared in the denominator, the mass of $\delta_{(3,2)}$ is not necessarily required to be heavy
 (the parameters $\lambda$ and $\mu$ are constrained to be small since they appear in the neutrino mass matrix). In this way one understands 
 the $B-L$ violating contribution to proton decay mediated by $\delta_{(3,2)}$. 

\begin{figure}[h]
\includegraphics[width=0.9\linewidth]{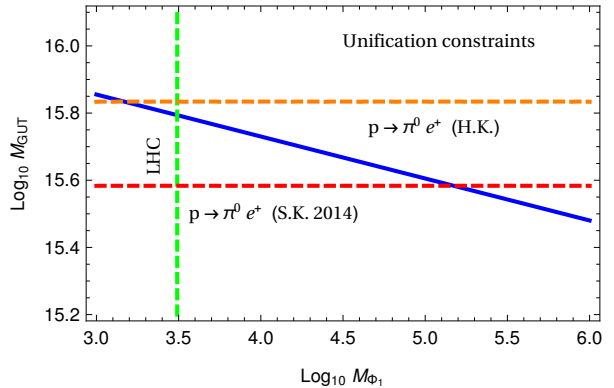}
\caption{Unification constraints shown by the blue line when $M_{H_2} = 1$ TeV. The dashed green line shows the naive LHC bound on the colored octet mass, $M_{\Phi_1}>3.1 $ TeV~\cite{Khachatryan:2015dcf}.
The red dashed line shows the limit on the GUT scale from the current experimental value on proton decay lifetime, $\tau_p (p \to \pi^0 e^+) > 1.29 \times 10^{34}$ years~\cite{Nishino:2012bnw}.
The orange dashed line shows the projected limit on the proton decay lifetime from the Hyper-Kamiokande collaboration, $\tau_p (p \to \pi^0 e^+) > 1.3 \times 10^{35}$ years~\cite{Miura}.
The mass of the $\Phi_3$ is in the range $10^{8.6}-10^{8.9}$ GeV from left to right.}
\end{figure}

In this model one has the usual doublet-triplet splitting problem since we need to split the $5_H$ representation and assume that the $T$ field is very heavy. Now, in the $45_H$ one has the same fine-tuning problem because the second Higgs $H_2$ must be light in order to have a large vacuum expectation value needed to correct the fermion masses. The simplest way to show that unification can be achieved in a consistent way is to assume only the splitting of the $45_H$ representation. 
We will assume that there is no mass splitting in $10_H$ or in $24_H$, and show the unification constraints in the scenarios where less fine-tuning is needed.

In Fig.~2, assuming unification of the gauge couplings at one-loop level, we show the allowed values for the masses of the $\Phi_1 \sim (8,2,1/2)$ and the unification scale. 
The solutions in agreement with the experiments are shown by the blue line. In this case we assume that $M_{H_2}=1$ TeV and 
the mass of $\Phi_3 \sim (3,3,-1/3)$ changes from $10^{8.6}$ and $10^{8.9}$ GeV. The couplings of the $\Phi_3$ to matter are not constrained by the fermion masses 
so that they can be small to suppress proton decay.
The green vertical line represents the LHC bound, $M_{\Phi_1} \geq 3.1$ TeV~\cite{Khachatryan:2015dcf}, 
on the colored octet mass. The red horizontal dashed line corresponds to the current experimental bounds on proton decay $\tau_p (p \to \pi^0 e^+) > 1.29 \times 10^{34}$ years~\cite{Nishino:2012bnw}.
We also show in Fig.~2 the limit projected (orange line) by the Hyper-Kamiokande collaboration, $\tau_p (p \to \pi^0 e^+) > 1.3 \times 10^{35}$ years~\cite{Miura}.
As one can appreciate, the main prediction from the unification of the gauge interactions is that the field $\Phi_1$ has to be light and the model could be tested in the near future in proton decay experiments.

In the case when the mass of $\Phi_1$ is close to the TeV scale one could hope to produce it at the LHC. The  Yukawa interactions for the field 
$\Phi_1$ are given by
\begin{equation}
 {\cal L}_Y \supset 2 d^c Y_2  \Phi_1^\dagger q_L \ + \ 4 u^c (Y_4-Y_4^T) q_L \Phi_1 \ + \ \textrm{h.c.}
 \label{Lphi1}
 \end{equation}
Notice that one can produce $\Phi_1$ with large cross sections through QCD interactions. 
The second term in the above equation comes from the interaction $10 \ 10 \ 45_H$.
Notice that since the second coupling above is anti-symmetric the decays into two top-quarks would not be observed.
Therefore, one can have exotic signatures such as signals with one top quark and three light jets. 
The phenomenological aspects of the colored octets have been investigated in Refs.~\cite{Manohar:2006ga,Gresham:2007ri,Gerbush:2007fe,Idilbi:2009cc,Kim:2009mha,Arnold:2009ay,Dorsner:2009mq,Ham:2010ha,Boughezal:2010ry,Idilbi:2010rs,Han,Arnold:2011ra,Bertolini:2013vta,Cao:2013wqa,He:2013tla,Khalil:2014gba,Fajer:2014ara,FileviezPerez:2008ib,Han:2010rf}.
%
\subsection{Proton Decay}
\label{sec:protondecay}
%
There are several fields mediating proton decay in this model. Here one has the usual gauge boson 
contributions mediated by the gauge bosons $V\sim (3,2,-5/6)$ living in $24_G$, and the Higgs contributions 
mediated by $T$ in $5_H$, as well as $\Phi_3, \Phi_5$ and $\Phi_6$ living in $45_H$. In the previous section we have investigated 
the unification constraints and we have shown that the only extra light Higgses are $H_2$ and $\Phi_1$ which do not 
mediate proton decay. Therefore, the main contribution to proton decay are the gauge contributions. 

The proton decay widths for the most relevant channels are given by
\begin{widetext}
\begin{eqnarray}
\Gamma (p \to \pi^0 e^+_\beta)&=& \frac{m_p}{8 \pi} A^2 k_1^4 \left(  \left| c(e^c,d) \matrixel{\pi^0}{(ud)_L u_R}{p}\right |^2  + \left| c(e,d^c) \matrixel{\pi^0}
{(ud)_R u_L}{p}\right|^2 \right), \\
\displaystyle \Gamma (p \to K^+ \bar{\nu}) &=& \frac{m_p}{8 \pi} \left( 1- \frac{m_{K^+}^2}{m_p^2}\right)^2 A^2 k_1^4 \sum_i  \left|  
c(\nu_i,d,s^c)  \matrixel{K^+}{(us)_R d_L}{p} + c(\nu_i,s,d^c)  \matrixel{K^+}{(ud)_R s_L}{p} \right|^2, \\
\textrm{with} \ A&=&A_{QCD} A_{SR}=\left( \frac{\alpha_3 (m_b)}{\alpha_3 (M_Z)} \right)^{6/23} \left( \frac{\alpha_3 (Q)}{\alpha_3 (m_b)} \right)^{6/25} \left( \frac{\alpha_3 (M_Z)}{\alpha_3 (M_{GUT})} \right)^{2/7}.
\end{eqnarray}
\end{widetext}
where $k_1=g_{GUT}/\sqrt{2} M_{GUT}$ and $A$ defines the running of the operators. $A_{QCD}\approx 1.2$ corresponds to the running from the $M_Z$ to the $Q\approx 2.3$ GeV scale, 
while $A_{SR} \approx 1.5$ defines the running from the GUT scale to the electroweak scale. The c-coefficients~\cite{FileviezPerez:2004hn} are given by
\begin{eqnarray}
&& c(e^c_\alpha, d_\beta) = V_1^{11} V_2^{\alpha \beta} + (V_1 V_{UD})^{1 \beta} (V_2 V_{UD}^\dagger)^{\alpha 1},\\
&& c(e_\alpha, d^c_\beta) = V_1^{11} V_3^{\beta \alpha},\\
&& c(\nu_l,d_{\alpha},d_{\beta}^c) = (V_1V_{UD})^{1\alpha}(V_3 V_{EN})^{\beta l}.
\end{eqnarray}
where the $V$'s are mixing matrices defined as 
\begin{eqnarray}
V_1=U_C^{\dagger}U, V_2=E_C^{\dagger}D, \  V_3=D_C^{\dagger}E, \\
V_{UD}=U^{\dagger}D \  \textrm{and} \  V_{EN}=E^{\dagger}N. 
\end{eqnarray}
The matrices $U$, $E$, $D$ and $N$ define the Yukawa couplings diagonalization so that 
\begin{eqnarray}
U_C^T Y_u U=Y_u^{\text{diag}}, \
D_C^T Y_d D=Y_d^{\text{diag}}, \\ 
E_C^T Y_e E=Y_e^{\text{diag}},\
N^T Y_\nu N=Y_\nu^{\text{diag}}.
\end{eqnarray}
The quantities $\matrixel{\pi^0}{(ud)_L u_L}{p}$, $\matrixel{\pi^0}{(ud)_R u_L}{p}$, $\matrixel{K^+}{(us)_R d_L}{p}$ and $\matrixel{K^+}{(ud)_R s_L}
{p}$ are the different matrix elements computed in lattice calculations.
Here we use the values reported in Ref.~\cite{Aoki}.
In general one cannot predict the c-coefficients entering in the decay width for the proton decay amplitude. In the most conservative scenario 
$c(e,d^c)= 1$, and $c(e^c, d) = 2$ for $p \to \pi^0 e^+$ and in the case of $p \to K^+ \bar{\nu}$ we use $c(\nu_l,d,s^c)= (V_3 V_{EN})^{2 l}$, 
and $c(\nu_l,s,d^c) = V_{CKM}^{12} (V_3 V_{EN})^{1 l}$. 

We show in Fig.~3 the conservative values for the proton decay lifetime and the current experimental bounds, $\tau_p (p \to \pi^0 e^+) > 1.29 \times 10^{34}$ years~\cite{Nishino:2012bnw} and $\tau_p ( p \to K^+ \bar{\nu}) > 5.9 \times 10^{33}$ years~\cite{SK:2014}. As one can see, in the Hyper-Kamiokande experiment one could rule out the model if proton decay is not found in the $\pi^0 e^+$ channel. In this way we show that the unification can be realized in agreement with the bounds on the proton decay lifetimes and the fact that the predictions are not far from the reach of 
the Hyper-Kamiokande experiment, one could hope to test this model in the near future.
\begin{figure}[h]
\includegraphics[width=0.9\linewidth]{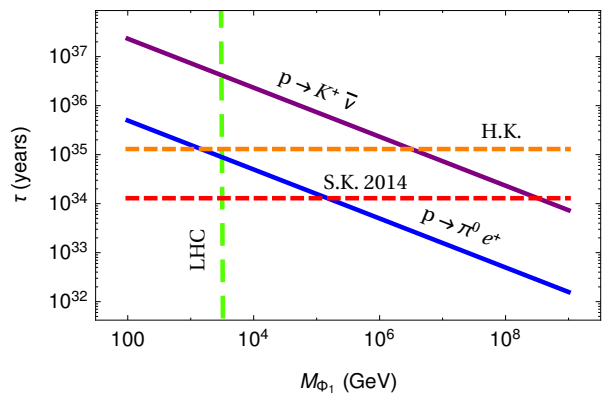}
\caption{Predictions for the proton decay lifetimes. The blue line shows the predictions for the decay $p \to \pi^0 e^+$, while the purple line shows the predictions for the decay $p \to K^+ \bar{\nu}$. 
The horizontal red dashed line shows the current experimental value on proton decay lifetime, $\tau_p (p \to \pi^0 e^+) > 1.29 \times 10^{34}$ years~\cite{Nishino:2012bnw} from the Super-Kamiokande collaboration.
The orange dashed line shows the projected limit on the proton decay lifetime from the Hyper-Kamiokande collaboration, $\tau_p (p \to \pi^0 e^+) > 1.3 \times 10^{35}$ years~\cite{Miura}. The green vertical line represents the LHC bound, $M_{\Phi_1} \geq 3.1$ TeV~\cite{Khachatryan:2015dcf}, 
on the colored octet mass.}
\end{figure}
\section{Summary}
We have discussed the simplest non-supersymmetric theories based on $SU(5)$ and 
pointed out that the minimal renormalizable theory without gauge singlets corresponds 
to the case where neutrino masses are generated at the quantum level through the Zee mechanism.
In this case the $45_H$ plays two major roles, it corrects the relation 
between charged lepton and down quark masses and generate neutrino masses through the Zee mechanism.

We have shown the possibility to have the unification of the gauge interactions in agreement with the experiments.
In most of the allowed parameter space the colored octet present in the theory is light so that it could give rise 
to exotic signals at the Large Hadron Collider. We have also investigated the predictions for proton decay showing the possibility 
to test this model at the Super-Kamiokande experiment or at the future Hyper-Kamiokande. The simple model proposed in this 
letter can be considered as one of the simplest renormalizable grand unified theories which motivates new experimental 
searches for proton decay and exotic signals at colliders. 

\textit{Acknowledgments}: The work of C.M. has been supported by the La Caixa-DAAD fellowship.

\section*{Appendix}
Here we show the different fields living in the $45_H$ and $10_H$ representations:
\begin{widetext}
\begin{eqnarray}
 \displaystyle 45_H&\sim&\underbrace{(8,2,1/2)}_{\Phi_1}\oplus \underbrace{(\bar{6},1,-1/3)}_{\Phi_2}\oplus \underbrace{(3,3,-1/3)}_{\Phi_3}
 \oplus \underbrace{(\bar{3},2,-7/6)}_{\Phi_4}\oplus \underbrace{(3,1,-1/3)}_{\Phi_5} \oplus \underbrace{(\bar{3},1,4/3)}_{\Phi_6}
 \oplus \underbrace{(1,2,1/2)}_{H_2} ,\\ \label{45H}
 \displaystyle 10_H& \sim&
 \underbrace{(1,1,1)}_{\delta^{+}}\oplus \underbrace{(3,2,1/6)}_{\delta_{(3,2)}}\oplus \underbrace{(\bar{3},1,-2/3)}_{\delta_{T}} \label{10H}.
\end{eqnarray}
\end{widetext}
The coefficients entering in the neutrino mass matrix are given by
\begin{eqnarray}
 c_e &=& \frac{(1 - 4 \sin^2 \beta)}{ 8 \pi^2 \sqrt{2} v \sin 2 \beta} \sin 2 \theta_+ \rm{Log} \left( \frac{m^2_{h_2^+}}{m^2_{h_1^+}} \right), \\
  c_d &=& \frac{1}{8 \pi^2 \sqrt{2}v \sin 2 \beta} \sin 2 \theta_+  \rm{Log} \left( \frac{m^2_{h_2^+}}{m^2_{h_1^+}} \right). 
\end{eqnarray}
\\
\newpage


  \end{document}